\documentstyle[preprint,aps,epsfig]{revtex}

\def\be{\begin{equation}}
\def\ee{\end{equation}}
\def\lk{\left(}
\def\rk{\right)}

\begin{document}
\draft

\title{New nonlinear dielectric  materials: Linear electrorheological fluids under the influence of electrostriction}


\author{J. P. Huang}
\address{
Department of Physics, The Chinese University of Hong Kong, Shatin, NT, Hong Kong, and
Max-Planck-Institut f\"ur Polymerforschung, Ackermannweg 10,
 55128, Mainz, Germany
%
}

\maketitle

\begin{abstract}

The usual approach to the development of new nonlinear  dielectric
materials focuses on the search for materials in which the
components possess an inherently large nonlinear dielectric response.
In contrast, based on thermodynamics, we have presented a
first-principles approach to obtain the electrostriction-induced
effective third-order nonlinear susceptibility  for the
electrorheological (ER) fluids in which the components have
inherent {\it linear}, rather than nonlinear, responses. In
detail, this kind of nonlinear susceptibility is in general of
about the same order of magnitude as the compressibility of the
linear ER fluid at constant pressure. Moreover, our approach has
been demonstrated in excellent agreement with a different
statistical method. Thus, such {\it linear} ER fluids  can serve as
a new nonlinear dielectric  material.


\end{abstract}



\newpage


Nonlinear   materials  with large values of the third-order
nonlinear susceptibility $\chi$~\cite{Smith84} are in great need
in industrial applications, such as, nonlinear optical switching
devices for use in photonics and real-time coherent optical signal
processors, and exploiting new type of nonlinear dielectric materials for use in  electronic and microwave components and sensor windows. 
It is usually believed that the effective nonlinear dielectric  
response can appear in the composite   material, in which at
least one component should possess an inherent {\it nonlinear}
response. Thus, the common way to  develop new nonlinear dielectric 
materials is to seek for materials in which the components possess
an inherently large nonlinear response~\cite{Dodenberger92}.  In
contrast, by using a first-principles approach based on
thermodynamics, we shall present a quite different method to
obtain the nonlinear dielectric  response from the electrorheological
(ER) fluids in which the components possess inherent {\it linear}
responses only (namely, linear ER fluids), under the influence of
electrostriction.  Thus, such linear ER fluids can also serve as
a  nonlinear dielectric  material because the effective third-order nonlinear susceptibility can be induced  due to the electrostriction effect. For clarity, it is worth noting that the above-mentioned linear ER fluids  represent the ER fluids whose dielectric constants are independent of the external electric field.

When an ER fluid~\cite{Winslow49,Tao91,Halsey92,Tao94,Sheng03} is
subjected to a strong external field, elongated chains or columns
of polarizable dielectric particles (e.g., titanium particles)
form immediately parallel to the field due to the anisotropic
long-range particle interaction inside the liquid carrier (e.g.,
silicone or corn oil). Because of this sort of rapid field-induced
aggregation, recently ER fluids have received much
attention~\cite{Tao91,Halsey92,Tao94,Sheng03} in both  scientific
research and industrial applications. For instance, ER fluids were also proposed as a method of constructing shock absorbers on
magnetically levitated trains.

Let us start by considering what electrostriction effect is for ER fluids.  In the presence of an inhomogeneous electric field ${\bf E},$ it is known that  a
translational force ${\bf F}_t$ exerts on a particle, which is
given by
 \be {\bf
F}_t = \alpha {\bf E}\cdot \nabla {\bf E}, \ee where $\alpha$
represents the polarizability of the particle. Therefore, an
inhomogeneous field acting on an ER fluid causes a particle concentration
gradient with high concentrations at high field strengths. Next,
if the ER fluid is situated partially in a strong external electric  field  at constant pressure, the density of the ER fluid in the field
will increase accordingly due to the interaction between the induced dipole
moment inside the particles and the electric field, which in turn yields an increase in the effective dielectric
constant. This effect is called electrostriction. In fact, the
phenomenon of electrostriction has been extensively studied, e.g.
for dipolar fluids~\cite{Bottcher73},  near-critical sulfur hexafluoride in miscrogravity~\cite{Zimmerli99},   ferroelectric liquid-crystalline elastomers~\cite{Lehmann01},   an all-organic composite consisting of polyvinylidene fluoride trifluoroethylene copolymer matrix and copper-phthalocyanine  particles~\cite{Li03}.
Regarding the ER system, one~\cite{Kim02} studied the electrostriction of solid ER composites  in an attempt to apply them in sensing shear stresses and strains in active damping of vibrations due to the high sensitivity of ER composites to shear electrostriction.
To the best of our knowledge, there is
neither theoretical nor experimental research  which treats the
electrostriction effect of ER fluids.
In this paper, based on thermodynamics we shall  present a
first-principles approach to derive  the
 electrostriction-induced effective nonlinear third-order
susceptibility $\chi$ of the linear ER fluids.




For investigating the electrostriction effect, take the experimental situation as follows: There is a capacitor with volume $V_c ,$ in which
the electric field and the dielectric displacement   are denoted
by $E_c$ and $D_c,$ respectively.  Both of them should satisfy the
usual electrostatic equations, namely
\begin{eqnarray}
\nabla\cdot {\bf D}_c &=& 0,\\
\nabla\times {\bf E}_c &=& 0.\label{NabE}
\end{eqnarray}
Here Eq.~(\ref{NabE}) implies that the electric field  ${\bf E}_c$
can be expressed as the gradient of a potential $\phi ,$ \be {\bf
E}_c=-\nabla \phi. \ee Under the appropriate boundary condition,
the inhomogeneous ER fluid (in the capacitor) can be represented
as a region of volume $V_c$, surrounded by surface $S_{{\rm s}}.$
Such kind of boundary condition is \be \phi = -{\bf E}\cdot {\bf
R}\,\,{\rm on}\,\,S_{s}, \ee which, if the ER fluid within $V_c$
were uniform, would give rise to an electric field which is
identical to ${\bf E}$ everywhere  within $V_c$. As a matter of
fact, even in an inhomogeneous ER fluid with this boundary
condition, the volume average of the electric field $\langle {\bf
E_c}\rangle$ within $V_c$ still equals that of the external field
$\langle{\bf E}\rangle$, i.e. \be \langle {\bf
E_c}\rangle=\frac{1}{V_c}\int {\bf E}_c({\bf R}){\rm d}^3r =
\langle {\bf E}\rangle \ee
It is worth noting that in this case there is no applied field
outside the capacitor. Also, the whole ER fluid with volume $V$ is
situated both inside and outside the capacitor at a constant
pressure $p.$


In the presence of the inhomogeneous external electric field ${\bf
E}$, the effective linear dielectric constant $\epsilon_e$ and
effective third-order nonlinear susceptibility $\chi$ for the ER
fluid inside the capacitor are defined as \be \langle {\bf
D}_c\rangle=\epsilon_e\langle {\bf E}\rangle+ 4\pi \chi |\langle {\bf
E}\rangle |^2\langle {\bf E}\rangle,\label{delf} \ee where
$\langle\cdots\rangle$ denotes the volume average of $\cdots .$ A
similar definition~\cite{Stroud88} was used for a composite
material which is subjected to a homogeneous external electric
field. In view of the real quantities under consideration,
Eq.~(\ref{delf}) can be rewritten as
 \be  \langle {\bf
D}_c\rangle = \epsilon_e\langle {\bf
E}\rangle+4\pi\chi \langle {\bf E}\rangle^2\langle {\bf E}\rangle.\label{chi-Def}
 \ee


On the other hand, based on thermodynamics the effective
dielectric constant $\epsilon_E$ including the incremental part
due to the electrostriction is defined as \be \epsilon_E
\equiv\lk\frac{\partial \langle {\bf D}_c\rangle}{\partial \langle
{\bf E}\rangle}\rk_{T,p}= \lk\frac{\partial \langle {\bf
D}_c\rangle}{\partial \langle {\bf E}\rangle}\rk _{T,\rho}+\int
f(d)\lk\frac{\partial \langle {\bf D}_c\rangle}{\partial\rho (d)}
\rk_{T,\langle {\bf E}\rangle}\lk\frac{\partial\rho (d)}{\partial
\langle {\bf E}\rangle} \rk_{T,p}{\rm d}d,\label{muE} \ee where
$\rho(d)$ stands for the density of the particles with diameter
$d,$ and $T$ temperature.   Here $\lk\frac{\partial \langle {\bf D}_c\rangle}{\partial
\langle {\bf E}\rangle}\rk _{T,\rho}$ corresponds to the
effective linear dielectric constant, namely $\epsilon_e .$
In Eq.~(\ref{muE}),  $f(d)$
  denotes a specific size distribution which exists in real ER fluids~\cite{Park98}, e.g. the lognormal
  distribution
$
f(d)=\frac{1}{\sqrt{2\pi}\sigma d}\exp [-\frac{\ln^2 (d/\delta)}{2\sigma^2}],\label{lnD}
$
where $\sigma$ is the standard deviation and $\delta$ the median diameter.

Accordingly, the incremental dielectric constant due to the
electrostriction [the last term of Eq.~(\ref{muE})] is equivalent to
$12\pi\chi\langle {\bf E}\rangle^2$. That is,
\be 12\pi\chi\langle {\bf E}\rangle^2=\int f(d)\lk\frac{\partial
\langle {\bf D}_c\rangle}{\partial\rho (d)} \rk_{T,\langle {\bf
E}\rangle}\lk\frac{\partial\rho (d)}{\partial \langle {\bf
E}\rangle} \rk_{T,p}{\rm d}d.\label{chiE2} \ee 
Let us take one
step forward to rewrite Eq.~(\ref{chiE2}) as \be \chi\langle {\bf
E}\rangle^2 = \frac{1}{12\pi} \int  f(d)\langle {\bf
E}\rangle\lk\frac{\partial\epsilon_e}{\partial\rho
(d)}\rk_{T,\langle {\bf E}\rangle}\lk\frac{\partial\rho
(d)}{\partial \langle {\bf E}\rangle}\rk_{T,p}{\rm d}d,\label{mue}
\ee The differential increase of the density inside the capacitor
${\rm d}\rho(d)$ corresponds to the increase in mass equal to
$V_c{\rm d}\rho(d).$ Naturally, this increase in mass is equal to
a decrease in mass outside the capacitor, which is given by
$-\rho(d) {\rm d}(V-V_c)=-\rho (d) {\rm d}V,$ so that ${\rm
d}\rho(d)=-[\rho(d)/V_c]{\rm d} V.$ Consequently, we may rewrite
Eq.~(\ref{mue}) as \be  \chi\langle {\bf E}\rangle^2 = -\frac{1}{12\pi}\int
f(d)\langle {\bf E}\rangle \frac{\rho(d)}{V_c}
\lk\frac{\partial\epsilon_e}{\partial\rho(d)}\rk_{T,\langle {\bf
E}\rangle}\lk\frac{\partial V}{\partial \langle {\bf
E}\rangle}\rk_{T,p}{\rm d}d.\label{mue2} \ee

Next, we can obtain $\lk\frac{\partial V}{\partial \langle {\bf
E}\rangle}\rk_{T,p}$ by using  the differential of the free energy
${\rm d}F $ \be {\rm d}F=-p{\rm
d}V-S{\rm d} T+\frac{V_c}{4\pi}\langle {\bf E}\rangle{\rm d}\langle {\bf
D}_c\rangle, \ee where $S$ denotes the entropy.
 In view of the transformed free enthalpy $G$
\be G = F+pV-\frac{V_c}{4\pi}\langle {\bf E}\rangle \langle
{\bf D}_c\rangle, \ee the differential of $G$ admits the form \be {\rm
d} G = -S{\rm d}T+V{\rm d}p-\frac{V_c}{4\pi}\langle {\bf
D}_c\rangle{\rm d}\langle {\bf E}\rangle. \ee
Based on this equation, we obtain
 \be \lk\frac{\partial V}{\partial \langle {\bf E}\rangle}
\rk_{T,p}  =
-\frac{V_c\langle {\bf
E}\rangle}{4\pi}\lk\frac{\partial\epsilon_e}{\partial p}
\rk_{T,\langle {\bf E}\rangle}.\label{differe}
 \ee 
Then, the substitution of Eq.~(\ref{differe}) to Eq.~(\ref{mue2}) yields \be  \chi\langle {\bf E}\rangle^2 =
\frac{1}{48\pi^2}\int  f(d)\langle {\bf
E}\rangle^2\rho(d)\lk\frac{\partial\epsilon_e}{\partial
\rho(d)}\rk_{T,\langle {\bf
E}\rangle}\lk\frac{\partial\epsilon_e}{\partial p}\rk_{T,\langle
{\bf E}\rangle}{\rm d}d.\label{muee} \ee 
Now let us use 
\be
\lk\frac{\partial\epsilon_e}{\partial p}\rk_{T,\langle {\bf
E}\rangle} =\beta\rho(d)\lk\frac{\partial\epsilon_e}{\partial\rho(d)}\rk_T,\label{compre}
\ee 
where \be\beta=-\frac{1}{V}\lk\frac{\partial V}{\partial
p}\rk_T\ee
 denotes the compressibility in the absence of the external electric field.  For deriving
 Eq.~(\ref{compre}), we have neglected the terms which depends on $\langle
{\bf E}\rangle$ because they lead to terms in
powers of $\langle {\bf E}\rangle$ higher than the second in
Eq.~(\ref{muee}). In the light of  the same approximation, the
substitution of Eq.~(\ref{compre}) into Eq.~(\ref{muee}) leads to \be
 \chi\langle {\bf E}\rangle^2 =\frac{1}{48\pi^2} \int  f(d)\langle
{\bf
E}\rangle^2\beta\rho(d)^2\lk\frac{\partial\epsilon_e}{\partial
\rho(d)}\rk^2 _T{\rm d}d. \ee
So far, the effective third-order nonlinear susceptibility $\chi$
of the ER fluid
is given by \be \chi = \frac{\beta}{48\pi^2}\int
f(d)\rho(d)^2\lk\frac{\partial\epsilon_e}{\partial
\rho(d)}\rk^2 _T{\rm d}d.\label{Xi} \ee
For determining the effective linear dielectric constant $\epsilon_e,$  we can resort to the
anisotropic Maxwell-Garnett theory, namely
\be \frac{g_L(\epsilon_e-\epsilon_2)}{\epsilon_2+g_L(\epsilon_e-\epsilon_2)}=
\frac{4\pi}{3}\int f(d)\frac{\rho(d)}{m(d)}\alpha (d){\rm d}d,
\label{epe} \ee where $m(d)$ ($\alpha(d)$) denotes the mass
(polarizability) of the individual particle with diameter $d,$
and $\epsilon_2$ the dielectric constant of the carrier liquid. It
is known that in the presence of an electric field, the particle
chain can be formed in the direction of the field, and thus the
structural anisotropy should appear inside this ER fluid.
Accordingly, in Eq.~(\ref{epe})
 $g_L$ ($g_L \ge 1/3 $) is the local field factor in the longitudinal field case, which was measured by using
 computer simulations~\cite{Martin98},  satisfying the sum rule $g_L+2g_T=1$~\cite{Huang01}.
 Here $g_T$ represents the local field factor in the transverse field case. As $g_L=1/3,$ the usual Clausius-Mossotti equation recovers, which is valid for an isotropic system.   In fact, the degree of
 anisotropy of the present system is  measured by how $g_L$ is deviated
 from $1/3.$  


Eq.~(\ref{Xi}) is the main result of the present paper. In detail,
the electrostriction-induced third-order nonlinear susceptibility
$\chi$ can be expressed in terms of the size distribution function and 
density of the particles,  the effective linear dielectric
constant, etc. In particular, it is apparent to see that at constant
pressure $\chi$ is proportional to the compressibility of the ER
fluids of interest. More precisely, $\chi$ is of about the same
order of magnitude as the compressibility, which can be readily
measured in experiments. Let us compute $\chi$ for a real example of an ER fluid by Klingenberg {\it et al.}~\cite{Klingenberg91}. In detail, this monodisperse ER fluid contains hollow silica spherical particles embedded in a corn oil. The parameters obtained from the experiment: diameter of the particles $95\,$$\mu m$,  $\epsilon_2=2.9$, dielectric constant of the particles $11$, volume fraction of the particles $0.26$,  appraent density $0.74\,$g/ml, and density of the corn oil $0.92\,$g/ml. Based on these parameters, it is straightforward to obtain $\alpha=1.5\times 10^{-7}\,$cm$^3$,  $\rho=0.228\,$g/cm$^3$, and $m=1.02\times 10^{-7}\,$g. If take $g_L=1/8$, $\chi=83.7\beta$. Further, as $\beta=2.1\times 10^{-10}\,$Pa$^{-1}$, $\chi=1.76\times 10^{-9}\,$(V/cm)$^{-2}$.
 To one's interest, Eq.~(\ref{Xi}) has exactly bridged the mechanical properties and 
nonlinear dielectric  properties of the linear ER fluids. In other
words, the
mechanical properties give rise to  the nonlinear dielectric responses (third-order nonlinear
susceptibilities) of the linear ER fluids.



In what follows, we would like to  show the correctness of the
present theory by comparing with a different statistical method.  First, let us
derive the increase of the density $\Delta\rho$ due to
electrostriction, based on   $\lk \partial V/\partial \langle {\bf
E}\rangle\rk_{T,p} .$ Let us start from \be \Delta\rho
= \int  \int_0^{\langle {\bf E}\rangle}
f(d)\lk\frac{\partial\rho(d)}{\partial \langle {\bf
E}\rangle}\rk_{T,p} {\rm d}\langle {\bf E}\rangle{\rm d}d. \ee
To this end, we  obtain \be \Delta\rho =\frac{1}{8\pi} \int f(d)
\langle {\bf E}\rangle^2\beta
\rho(d)^2\lk\frac{\partial\epsilon_e}{\partial \rho(d)}\rk_T {\rm
d}d. \label{delta}\ee 
Again, in the expression for $\Delta\rho$ terms in powers
of $\langle {\bf E}\rangle$ higher than the second have been
neglected. For a monodisperse case, Eq.~(\ref{delta}) reduces to \be \Delta \rho=
\frac{1}{8\pi}  \langle {\bf E}\rangle^2\beta
\rho^2\lk\frac{\partial\epsilon_e}{\partial \rho}\rk_T. \ee

Let us assume there is an ideal gas inside the capacitor. In this
case, the compressibility is given by 
\be 
\beta =\frac{M}{\rho
RT}, 
\ee 
where $M$ is the molecular weight, and $R$ the molar gas constant. For the ideal gas
(monodisperse case), setting $g_L=1/3$ to the above
Clausius-Mossotti equation [Eq.~(\ref{epe})] yields \be
\frac{\epsilon_e-1}{\epsilon_e+2}=\frac{4\pi}{3}\frac{\rho}{m}\alpha
.\ee In view of $\epsilon_e-1\ll 1$ for ideal gases, we obtain \be
\lk \frac{\partial\epsilon_e}{\partial\rho}\rk_T =\frac{4\pi
}{m}\alpha ,\ee and hence  the desired results for $\Delta\rho ,$
\be
\Delta\rho=\frac{\langle {\bf E}\rangle^2\rho\alpha}{2k_BT}.\label{eq1}\ee


This equation can also be achieved by using a  statistical
method. According to Boltzmann's distribution
law, the number of moles per cm$^3$ of the gas at a point with
field strength $\langle {\bf E}\rangle$ is given by \be N = N'
\exp (-\frac{W}{k_BT}), \ee where $W$ denotes the average value of
the work required to bring a molecule into the field $\langle {\bf
E}\rangle ,$ and $N'$ the number of moles per cm$^3$ of the gas at
a point in the absence of field. It is straightforward to obtain
\be
\Delta\rho = M (N-N')=\frac{\langle {\bf E}\rangle^2\rho\alpha}{2k_BT},\label{eq2}
\ee
which is exactly the same as Eq.~(\ref{eq1}). Again, the terms in higher powers of $\langle {\bf E}\rangle$ than the second have been neglected.

To sum up, by using thermodynamics we have presented a
first-principles approach to the derivation of the effective
third-order nonlinear susceptibility  [Eq.~(\ref{Xi})] of the
linear ER fluids under the influence of electrostriction, which is
of about the same order of magnitude as the compressibility of the
 ER fluid at constant pressure. Our approach has been demonstrated in excellent agreement with
an alternative statistical method.



The aim of the present paper is to exploit electrostriction in a linear ER fluid in order to generate a nonlinear dielectric response. The proposed mechanism works for dc electric fields. It should also be expected to work for ac fields 
with  frequency $\nu$ if the size of the sample 
is not greater than $c_{s}/\nu$, where $c_{s}$ is the sound velocity. In this connection, 
$\nu$ can be up to kHz or so. Otherwise the required mass density oscillations 
will not be able to keep up with the rapid changes in the electric field. 

The theory described in this paper can be used to study any
colloidal suspensions  like magnetorheological
fluids~\cite{Kordonsky91}, ferrofluids~\cite{Rosensweig85}, et al.
Since there exist permanent magnetic dipole moments inside the
magnetorheological fluids and  ferrofluids, the derivation of
the effective linear permeability can still be done by using the present anisotropic
Clausius-Mossotti equation, in which, however, the terms of
permanent magnetic moments should be added accordingly. In a word,
we have shown theoretically that the linear ER fluids under the
influence of the electrostriction effect can serve as a new
nonlinear dielectric  material.

\section*{Acknowledgments}

This work was in part supported  by the Alexander von Humboldt Foundation of Germany. The author acknowledges Professor K. W. Yu's fruitful discussions.

 \newpage


\newpage

\begin{references}

\bibitem{Smith84} See, for example, P. W. Smith, Philos. Trans. R. Soc. London A {\bf
313}, 349 (1984); {\it Nonlinear Photonics}, edited by H. M.
Gibbs, G. Khitrova, and N. Peyghambarian (Springer-Verlag, New
York, 1990); G. I. Stegeman, in {\it Contemporary Nonlinear
Optics}, edited by G. P. Agrawal and R. W. Boyd (Academic Press,
Boston, 1992).

\bibitem{Dodenberger92} See, for example, D. C. Dodenberger, J. R.
Heflin, and A. F. Garito, Nature (London) {\bf 359}, 309 (1992);
D. J. Bergman and D. Stroud, Solid State Phys. {\bf 46}, 147 (1992); G. L. Fischer, R. W. Boyd, R. J. Gehr, S. A. Jenekhe,
 J. A. Osaheni, J. E. Sipe and L. A. Weller-Brophy,
 Phys. Rev. Lett. {\bf 74}, 1871 (1995).


\bibitem{Winslow49} W. M. Winslow, J. Appl. Phys.{\bf 20}, 1137 (1949).

\bibitem{Tao91} R. Tao, J. M. Sun, Phys. Rev. Lett. {\bf 67}, 398 (1991).

\bibitem{Halsey92} T. C. Halsey, Science {\bf 258}, 761 (1992).

\bibitem{Tao94} For example, see {\it Electrorheological Fluids}, edited by R. Tao (World Scientific, Singapore, 1992);
 {\it Electrorheological Fluids}, edited by R. Tao and G. D. Roy (World Scientific, Singapore, 1994);
 {\it Electro-Rheological Fluids, Magneto-Rheological Suspensions and Associated Technology},
 edited by W. A. Bullough (World Scientific, Singapore, 1996);
  {\it Electrorheological Fluids and Magetorheological Suspensions}, edited by G. Bossis
  (World Scientific, Singapore, 2001).

\bibitem{Sheng03} W. J. Wen, X. X. Huang, S. H. Yang, K. Q. Lu,
and P. Sheng, Nature Materials {\bf 2}, 727 (2003).

\bibitem{Bottcher73}
C.\,J.\,F. B\"{o}ttcher, {\em Theory of electric polarization}, second edition,
(Elsevier, Amsterdam, 1993), Vol.~1.

\bibitem{Zimmerli99} G. A. Zimmerli, R. A. Wilkinson, R. A. Ferrell, and M. R. Moldover, Phys. Rev. Lett. {\bf 82}, 5253 (1999).

\bibitem{Lehmann01} W.  Lehmann, H. Skupin, C. Tolksdorf, E. Gebhard, R. Zentel, P. Kruger, M. Losche, F. Kremer, Nature (London) {\bf 410}, 447 (2001).

\bibitem{Li03} J. Y. Li, Phys. Rev. Lett. {\bf 90}, 217601 (2003).

\bibitem{Kim02} G. H. Kim and Y. M. shkel, J. Intel. Mat. Syst. Str. {\bf 13}, 479 (2002).

\bibitem{Stroud88} D. Stroud and P. M. Hui, Phys. Rev. B {\bf 37}, 8719 (1988).

\bibitem{Park98} C. Park and R. E. Robertson, Mat. Sci. Eng. A-Struct. {\bf 257}, 295 (1998); A. Kawai, K. Uchida, and F. Ikazaki, in  {\it Electrorheological Fluids and Magetorheological Suspensions}, edited by G. Bossis
  (World Scientific, Singapore, 2001), pp.~626-632.




\bibitem{Martin98} J. E. Martin, R. A. Anderson, and C. P. Tigges, J. Chem. Phys. {\bf 108}, 3765 (1998); ibid, {\bf 108}, 7887 (1998).



\bibitem{Huang01} J. P. Huang, J. T. K. Wan, C. K. Lo, and K. W. Yu, Phys. Rev. E {\bf 64}, 061505(R) (2001).

\bibitem{Klingenberg91} D. J. Klingenberg, F. V. Swol, and C. F. Zukoski, J. Chem. Phys. {\bf 94}, 6170 (1991)

\bibitem{Kordonsky91} V. I. Kordonsky and Z. P. Shulman, in {\it Electrorheological Fluids}, edited by J. D. Carlson, A. F. Sprecher, and H. Conrad (Technomic Publishing, Lancaster, Basel, 1991), pp.~437-444.

\bibitem{Rosensweig85} R. E. Rosensweig, {\it Ferrohydrodynamics} (Cambridge Univ. Press, Cambridge, 1985).
\end{references}
\end{document}